\documentstyle[aps,titlepage,preprint]{revtex}

\begin{document}

\title{
Non-Markovian quantum kinetics and conservation laws}
\author{V. G. Morozov\thanks{Permanent address:
Department of Physics, Moscow Institute
of Radioengineering,
Electronics and Automation,
Vernadsky Prospect, 78, 117 454, Moscow, Russia.}
\,\, and Gerd R{\"o}pke\\
Department of Physics, Rostock University, 18051 Rostock,
Germany\\
E-mail: gerd@darss.mpg.uni-rostock.de}
\maketitle

\begin{abstract}
A link between memory effects  
in quantum kinetic equations and nonequilibrium correlations
associated with the energy conservation
is investigated. In order that the energy be conserved by an
approximate collision integral,  
the one-particle distribution function and the
mean interaction energy are treated as independent
nonequilibrium state parameters.  
The density operator method is used to derive a 
kinetic equation in second-order non-Markovian Born approximation
and an evolution equation for the nonequilibrium 
quasi-temperature which is
thermodynamically conjugated to the mean interaction energy.
The kinetic equation contains a correlation contribution 
which exactly cancels the collision term in thermal equilibrium 
and ensures the energy conservation in nonequilibrium states.
Explicit expressions for the entropy production
in the non-Markovian regime and the time-dependent correlation
energy are obtained. 
\end{abstract}

\vspace*{10pt}

 
\setcounter{equation}{0}

\section{Introduction}
Experimental studies of fast relaxation processes
caused by the interaction of short laser pulses with
matter~\cite{Shan96,HaugJauho96} have inspired a renewed
interest in non-Markovian kinetics. 
Although this subject has been under development for many
years, recent investigations have shown that
the inclusion of memory effects in collision integrals 
leads to some serious problems, which did not receive proper 
attention previously. 

Since in many cases of experimental interest the system
can be described in terms of weakly interacting quasiparticles, 
the most-used non-Markovian quantum kinetic equations are in
fact modifications of the 
so-called {\em Levinson equation\/}~\cite{Levinson69} 
in which the collision integral is taken in second-order Born
approximation and the energy-conserving delta function is 
replaced by an oscillating memory kernel
(see, e.g.~\cite{HaugJauho96}). 
It should be noted, however,
that such kinetic equations have some grave disadvantages.
(i) The Levinson-like kinetic equations have unstable solutions
and even produce negative distribution functions. 
(ii) If the initial state of the  system is already thermal
equilibrium, the collision integrals do not 
vanish, giving rise to non-physical time evolution.          
The first defect can be overcome by 
using certain decaying  memory 
kernels~\cite{HaugJauho96,Banyai92,HaugBanyai96}
which take account of the quasiparticle damping.
However, in such approaches, the problem of the 
equilibrium solution still persists because
{\em all\/} ``improved'' memory
kernels lead to collision integrals which do not conserve
the total energy of the system. As a consequence, in the  
long-time (Boltzmann) limit the ``improved''
non-Markovian quantum kinetic equations
do not describe the relaxation to thermal equilibrium, leading
to an overpopulation of high-energy 
states.  

To summarize, it appears that the quasiparticle damping
alone cannot be responsible for the long-time  
asymptotic behavior of
non-Markovian quantum kinetic equations; 
there must be another physical mechanism which provides the exact
cancellation of collision effects in thermal equilibrium
and does not violate the energy conservation.    
The origin of such a mechanism can be deduced from consideration
of kinetic processes in the presence of initial
correlations. As early as 1970 Lee 
{\em et al.}~\cite{LeeFujita70} studied the 
evolution of a weakly interacting low-density classical 
gas with a correlated initial state and showed that in 
equilibrium the changes in
the one-particle distribution function due to collisions
(including memory effects) and initial correlations exactly
cancel each other. 
Recently the same result was obtained
for quantum systems~\cite{MorozovRoepke99}. 
It therefore would appear reasonable that the 
interplay between collisions
and {\em nonequilibrium\/} correlations would
be the mechanism which provides the correct
long-time behavior of non-Markovian
quantum kinetic equations. 

This paper presents an approach 
in which the time evolution of the one-particle
distribution function is coupled with the evolution 
of long-lived correlations associated with  
conservation laws. 
The basic idea is to 
treat conserved quantities, 
most notably the energy,
as independent state parameters in addition to 
the one-particle distribution
function. Although this is not to say that {\em all\/} 
many-particle correlations can be incorporated in this way,  
the advantage of the above idea, originally suggested
in the context of a generalization of the 
Enskog theory to dense classical  gases with 
``soft'' inter-particle 
potentials~\cite{ZubMor84,VanBeijeren88} and then
applied to quantum systems~\cite{Loss90,MorozovRoepke95}, 
lies in the fact that now, due to microscopic equations of motion, 
the energy is an exactly conserved quantity in any approximation
for the collision integral. 
Moreover, the energy conservation leads to the appearance of 
additional correlation terms in the kinetic equation, which
substantially compensate the collision contribution. 
In this paper, within second-order
non-Markovian Born approximation,
we derive a collision integral involving the collision 
and correlation terms.  
It is shown that the kinetic equation is consistent with the
energy conservation and has the correct equilibrium solution.  
   
The paper is organized as follows.
In Section~II we consider a generalized Gibbs ensemble 
in which the energy and the one-particle distribution 
function play the role of given nonequilibrium state parameters. 
The corresponding {\em relevant statistical operator\/}
is used  to define the thermodynamic quantities: 
nonequilibrium entropy, the
quasi-temperature, and the quasi-chemical
potential. 
We also derive the  evolution equations for the thermodynamic
quantities  in terms of the collision integral.
Section~III is concerned with construction of the nonequilibrium
statistical operator describing non-Markovian kinetic processes 
in a spatially homogeneous system.  
The nonequilibrium statistical operator is 
found in terms of the relevant statistical operator for 
quantum systems with a weak
interaction by employing iterative solution of the von Neumann
equation. 
In Section~IV the nonequilibrium statistical operator is used to 
calculate the non-Markovian collision integral. 
In Section~V we derive the entropy production in the
non-Markovian regime and an expression for the time-dependent
correlation energy of the system. 
Section~VI sketches a generalization of the theory to
spatially non-homogeneous systems.  
Finally, Section~VII contains the final conclusions and comments on
further applications of the theory.

\setcounter{equation}{0}

\section{Nonequilibrium correlations associated with
conservation laws}
To put our discussion into a straightforward language,
we shall consider a system of fermions or bosons with
the Hamiltonian $\hat H=\hat H_0 +\hat H'$,
where $\hat H^{}_0$ is the kinetic energy operator
and the term $\hat H'$ describes a pairwise interaction
between particles. In second quantized form, these operators 
are given by 
\begin{equation}
\label{2.1}
\hat H_0=\sum_1 \varepsilon^{}_{11'}\,a^{\dagger}_{1} a^{}_{1'},
\qquad
\hat H'=
{1\over2}\sum_{121'2'} \langle 1'2'|V|12\rangle \,
a^{\dagger}_{2'}a^{\dagger}_{1'}a^{}_{1}a^{}_{2},
\end{equation}
where  the label $k$ denotes a complete
set of single-particle quantum numbers,
$\varepsilon^{}_{11'}$ is a hermitian single-particle
energy matrix, 
$a^{}_k$ and $a^{\dagger}_k$ are Fermi or Bose annihilation 
and creation operators. Generally speaking, 
the Hamiltonian can also contain
additional terms describing interaction of the system with
external fields. For simplicity, we will not consider this
interaction explicitly and 
restrict ourselves to relaxation processes in 
the system just after   
the initial excitation by the external field. 
However, the theory can easily be generalized to the case where
the field effects are taken into consideration 
(see Section~VII for a discussion). 

In what follows the system is assumed to be spatially
homogeneous, which is adequate for most experimental
situations where memory effects are 
of crucial importance~\cite{HaugJauho96}. 
Some aspects of the following analysis 
may also be of interest for transport processes in 
non-homogeneous systems, so that we shall return to this point in 
Section~VI. In the spatially homogeneous case, 
it is convenient to take $(k)=({\bf p}^{}_k,\sigma^{}_k)$,
where ${\bf p}$ is the momentum and
$\sigma$ is the spin index. 
Then the kinetic energy operator
in Eq.~(\ref{2.1}) becomes
\begin{equation}
\label{2.1a}
\hat H_0=\sum_{1} 
\varepsilon^{}_{1}\,a^{\dagger}_{1} a^{}_{1}
\end{equation}
where $\varepsilon_{1}=\varepsilon^{}_{p}$ 
are single-particle energies.

Within the kinetic description of the system, 
the main objective is to derive a kinetic equation for 
the one-particle distribution function
\begin{equation}
\label{2.3}
f^{}_1(t)=\langle a^{\dagger}_1 a^{}_1 \rangle^t
\equiv \langle \hat f^{}_1\rangle^t,  
\end{equation}
where the symbol $\langle \cdots\rangle^t$ stands for the average
calculated with some nonequilibrium statistical 
operator $\varrho(t)$.
Formally, the kinetic equation follows immediately from the von
Neumann equation for the statistical operator
\begin{equation}
\label{2.4}
\frac{\partial\varrho(t)}{\partial t}+
{1\over i\hbar}\left[\varrho(t),\hat H\right]=0.
\end{equation} 
Taking the product of this equation with 
$\hat f^{}_1$ and then calculating
the trace, we get
\begin{equation}
\label{2.5}
\frac{\partial f^{}_1(t)}{\partial t}= I^{}_1(t),
\end{equation}
where
\begin{equation}
I^{}_{1}(t)={1\over i\hbar}\,\langle [\hat f^{}_1,\hat H']\rangle^t=
{1\over i\hbar}\,{\rm Tr}\left\{[\hat f^{}_1,\hat H']\,
\varrho(t)\right\} 
\label{2.6}
\end{equation}
is the collision integral. 
In order that Eq.~(\ref{2.5}) be a closed
kinetic equation, the collision integral must be expressed in
terms of the one-particle distribution function --- which is
equivalent to requiring that the statistical operator 
$\varrho(t)$ is represented as a functional of the one-particle
distribution function. 
A common technique for obtaining such a representation is 
based on the condition of complete weakening of initial 
correlations for the quantum BBGKY hierarchy in combination with
some truncation procedure. A limitation of this standard scheme 
is the very strong assumption that the one-particle distribution
function is the only nonequilibrium state variable describing
the system. 
However, there exist long-lived correlations which
cannot be expressed {\em exactly\/} in terms of the one-particle
distribution function. For example, 
many-particle correlations arise due to formation of bound 
states.    
Another origin of long-lived correlations lies in local
conservation laws. The conservation of energy is of
special importance because the density of the interaction energy 
is determined by the two-particle distribution
function. 
Thus, strictly speaking, kinetic processes must always be
considered together with the evolution of locally conserved
quantities, i.e., with hydrodynamic processes. 
A consistent description of kinetics and hydrodynamics 
can be developed by treating the
one-particle distribution function 
(for quantum systems, the Wigner function) and the energy
density as independent state 
parameters, which means that the statistical operator
$\varrho(t)$ is represented as a functional of the corresponding
dynamical variables~(see, 
e.g.,~\cite{MorozovRoepke95,ZubMorRoep1,Tokarchuk98} and
references therein).
Here we will follow this approach to study
non-Markovian kinetic processes in a spatially homogeneous
system. In this case the correlation effects arise due to the
fact that the total energy is an integral of motion.
Modifications of the theory needed for the spatially
non-homogeneous case will be discussed in Section~VI.
                    
\subsection{The relevant statistical operator}
We begin by considering the statistical thermodynamics 
description of nonequilibrium states
with correlations caused by conservation laws.  
For a one-component spatially homogeneous system,
the conserved quantities of interest are the total number 
of particles, $N$,
the total momentum, ${\bf P}$, and the total energy, 
${\cal E}$.
They can be expressed as the mean values of the
dynamical variables 
$\hat N=\sum_1 \hat f^{}_1$,\, 
$\hat{\bf P}=\sum_1 {\bf p}^{}_1 \hat f^{}_1$,\, and
$\hat H$, i.e.,
\begin{equation}
\label{2.7}
N=\langle \hat N\rangle^t= \sum_1 f^{}_1(t),
\qquad
{\bf P}=\langle \hat{\bf P}\rangle^t= 
\sum_1 {\bf p}^{}_1  f^{}_1(t),
\qquad
{\cal E}=\langle \hat H\rangle^t.
\end{equation} 
For simplicity, we  shall assume  the total 
momentum  to be zero. Then the only integrals of motion that must 
be considered are the energy and the number of particles.

To see how the statistical thermodynamics description of 
nonequilibrium correlations can be developed, 
we recall the well-known 
grand canonical statistical operator
\begin{equation}
 \label{x2.1}
 \varrho^{}_{\rm eq}=
\exp\left\{ -\beta(\hat H -\mu \hat N)\right\}\left/
{\rm Tr}\,
\exp\left\{ -\beta(\hat H -\mu \hat N)\right\}
\right., 
\end{equation}
where the equilibrium inverse temperature, $\beta=1/T$,  
and the chemical potential, $\mu$,  are  related to 
$N$ and ${\cal E} $ by the equations of state
\begin{equation}
 \label{x2.2}
 N={\rm Tr}\left\{\hat N \varrho^{}_{\rm eq}\right\},
\qquad
{\cal E}= {\rm Tr}\left\{\hat H\varrho^{}_{\rm eq}\right\}. 
\end{equation}
The grand canonical operator~(\ref{x2.1}) describes the
equilibrium Gibbs ensemble
and is a stationary solution of the von Neumann 
equation~(\ref{2.4}). 
Suppose now that  we want to construct a statistical operator
which describes a generalized Gibbs ensemble characterized by
a {\em nonequilibrium\/} one-particle distribution function
$f^{}_1(t)$ and by given values of the conserved quantities.
This is a special case of a more general situation where 
the state of the system is described by a set  of parameters
$P^{}_m(t)$  which can be represented as the mean values,  
$P^{}_m(t)={\rm Tr}\left\{ \hat P^{}_m \varrho(t) \right\}$,
of some  dynamical variables $\hat P^{}_m$. 
As argued by Jaynes~\cite{Jaynes57}, the corresponding 
{\em relevant statistical operator\/} can be derived by 
maximizing the entropy functional for given
$P^{}_m(t)$. The extremum
condition for the entropy gives~\cite{ZubMorRoep1}
\begin{equation}
\label{e2.1}
\varrho^{}_{\rm rel}(t)=
\exp\left\{-\sum_{m} \lambda^{}_m(t) \hat P^{}_{m}\right\}\left/
{\rm Tr}\,\exp\left\{-\sum_{m} \lambda^{}_m(t) \hat P^{}_m
\right\} .
\right.
\end{equation}
The Lagrange multipliers $\lambda^{}_m(t)$ are to be expressed in
terms of the $P^{}_n(t)$ from the self-consistency conditions
\begin{equation}
\label{e2.3} 
P^{}_m(t)= {\rm Tr}\left\{\hat P_m\,
\varrho^{}_{\rm rel}(t)
\right\},
\end{equation}
which can also be interpreted  as nonequilibrium equations of state.
Following the above line of reasoning, it is easy to 
derive the relevant
statistical operator which describes the generalized 
Gibbs ensemble with 
given values of the  total energy, the total number of 
particles, and 
the one-particle distribution function. 
We write it in the form
\begin{equation} 
\label{2.13}
\varrho^{}_{\rm rel}(t)={1\over Z^{}_{\rm rel}(t)}\,
\exp\left\{-\beta^*(t)\left(\hat H -\mu^*(t)\,\hat N\right) 
- \sum_{1} \lambda^{}_1(t)\,\hat f^{}_1 \right\},
\end{equation}
where the partition function  is determined by the
normalization condition,
\begin{equation}
\label{2.14}
Z^{}_{\rm rel}(t)={\rm Tr}\,
\exp\left\{-\beta^*(t)\left(\hat H -\mu^*(t)\,\hat N\right)  
- \sum_{1} \lambda^{}_1(t)\,\hat f^{}_1 \right\},
\end{equation}
and the Lagrange multipliers  $\beta^*(t)$, $\mu^*(t)$, and
$\lambda^{}_1(t)$  are to be calculated from the self-consistency 
conditions (nonequilibrium equations of state)
\begin{equation}
\label{2.15}
f^{}_1(t)={\rm Tr}\left\{\hat f^{}_1\varrho^{}_{\rm rel}(t)\right\},
\qquad
N={\rm Tr}\left\{\hat N\varrho^{}_{\rm rel}(t)\right\},
\qquad
{\cal E}= {\rm Tr}\left\{\hat H\varrho^{}_{\rm rel}(t)\right\}.
\end{equation}
Note that the  grand canonical
operator~(\ref{x2.1})  is a special case
of  the relevant statistical operator~(\ref{2.13}), 
where  the state parameters are  integrals of motion.
 
\subsection{Alternative representations for the 
relevant statistical operator}
By analogy with  Eq.~(\ref{x2.1}), the quantities 
$T^*(t)=1/\beta^*(t)$
and $\mu^*(t)$ may be called the {\em quasi-temperature\/} and the
{\em quasi-chemical potential}, respectively. 
Here one comment  is in order. 
The important difference between the state parameters
$N$ and ${\cal E}$ is that the former is a linear combination of
the one-particle distribution functions, $f^{}_1(t)$,
whereas ${\cal E}$ involves the mean interaction energy
\begin{equation}
\label{2.8}
{\cal E}^{}_{\rm int}(t)=\langle\hat H'\rangle^t,
\end{equation}
which, in general, cannot be expressed in terms of $f^{}_1(t)$
only.  Thus, in fact, the set of independent state parameters
consists of $f^{}_1(t)$ and ${\cal E}^{}_{\rm int}(t)$.
This also is seen from Eq.~(\ref{2.13}) where
the terms coming from the 
kinetic energy operator and the particle number operator 
can be combined with the last term by introducing 
new Lagrange multipliers  $\Lambda^{}_1(t)$ through the
relation
\begin{equation}
\label{2.12}
\Lambda^{}_1(t)=\beta^*(t)\left[
\varepsilon^{}_1 - \mu^*(t) \right]
+\lambda^{}_1(t).  
\end{equation} 
Then Eq.~(\ref{2.13}) reduces to
\begin{equation}
\label{2.9}
\varrho^{}_{\rm rel}(t)={1\over Z^{}_{\rm rel}(t)}\,
\exp\left\{-\beta^*(t)\,\hat H' 
- \sum_{1} \Lambda^{}_1(t)\,\hat f^{}_1 \right\}
\end{equation}
with the partition function given by
\begin{equation}
\label{2.10}
Z^{}_{\rm rel}(t)=
{\rm Tr}\,\exp\left\{-\beta^*(t)\,\hat H' 
- \sum_{1} \Lambda^{}_1(t)\,\hat f^{}_1 \right\}.
\end{equation}
The self-consistency conditions for the Lagrange
multipliers $\beta^*(t)$  and $\Lambda^{}_1(t)$ can now 
be taken in the form
\begin{equation}
\label{2.11}
f^{}_1(t)={\rm Tr}\left\{\hat f^{}_1\varrho^{}_{\rm rel}(t)\right\},
\qquad
{\cal E}^{}_{\rm int}(t)
={\rm Tr}\left\{\hat H'\varrho^{}_{\rm rel}(t)\right\},
\end{equation}
since the mean kinetic energy is exactly expressed in terms of
the one-particle distribution function.
Summarizing, the quasi-chemical potential, $\mu^*(t)$, can, in
principle, be excluded from the set of Lagrange parameters since the
second of the conditions~(\ref{2.15}) follows from the first one. 
Nevertheless, 
the expression~(\ref{2.13}) has two advantages.
First, it goes over explicitly  to the equilibrium statistical 
operator, if
$\beta^*(t)=1/T$, $\mu^*(t)=\mu$, and $\lambda^{}_1(t)=0$.
Second, some  formulas  to be derived in the following  
have a more clear physical interpretation when written in terms of the 
quasi-chemical potential. For these reasons, we will consider 
Eq.~(\ref{2.13}) as a representation for $\varrho^{}_{\rm rel}(t)$
which is equivalent to Eq.~(\ref{2.9}) 
by virtue of  Eq.~(\ref{2.12}).
Note, however,  that Eq.~(\ref{2.12}) determines only 
$\lambda^{}_1(t)-\beta^*(t)\mu^*(t)$, but not the
quasi-chemical  potential itself.  Since  no physical quantity 
depends on the special choice of  the quasi-chemical potential, 
the function $\mu^*(t)$ may be chosen arbitrarily provided 
that $\mu^*(t)=\mu$ in thermal equilibrium.
For our purposes,  it will be convenient 
to define the quasi-chemical potential through the 
condition
\begin{equation}
\label{2.21}
N={\rm Tr}\left\{\hat N\varrho^{}_{\rm q}(t)\right\}, 
\end{equation}
where $\varrho^{}_{\rm q}(t)$ is the auxiliary
{\em quasi-equilibrium statistical operator}
\begin{equation}
\label{2.22}
\varrho^{}_{\rm q}(t)=
\exp\left\{-\beta^*(t)\left(\hat H -\mu^*(t)\,\hat N\right)\right\}
\left/
{\rm Tr}\,
\exp\left\{-\beta^*(t)\left(\hat H -\mu^*(t)\,\hat N\right)\right\},
\right.
\end{equation}
which describes the state characterized by
the quasi-temperature $T^*(t)=1/\beta^*(t)$ and the 
total number of particles $N$.
Equation~(\ref{2.21})
ensures that, in the equilibrium limit, $\mu^*(t)$ 
goes to the chemical potential $\mu$ since
$\beta^*(t)$ goes to the equilibrium inverse temperature.

For completeness, we give one more representation for the
relevant statistical operator, which is obtained from 
Eq.~(\ref{2.13})
by the formal decomposition of the Hamiltonian
\begin{equation}
\label{4.1}
\hat H=\hat{\cal H}^{}_0(t) +\hat{\cal H}'(t),
\end{equation}
where
\begin{equation}
\label{4.2}
\hat{\cal H}^{}_0(t)=\sum^{}_{1} E^{}_1(t)\,a^{\dagger}_1 a^{}_1,
\qquad
\hat{\cal H}'(t)=
\hat H' - \sum_{1} \Sigma^{\rm HF}_1(t)\,a^{\dagger}_1 a^{}_1,
\end{equation}
and the re-normalized  single-particle energies,
\begin{equation}
\label{3.7}
E^{}_1(t)=\varepsilon^{}_1 +\Sigma^{\rm HF}_1(t),
\end{equation}
involve the exchange Hartree-Fock term
\begin{equation}
\label{3.8}
\Sigma^{\rm HF}_1(t)= \sum_{2} \langle 12|V|12\rangle^{}_{\rm ex}\,
f^{}_2(t).
\end{equation}  
Here and in what follows the subscript  ``ex'' indicates the 
symmetrized (antisymmetrized) interaction amplitude
\begin{equation}
\label{3.5}
\langle 12|V|1'2'\rangle^{}_{\rm ex}=
\langle 12|V|1'2'\rangle
\mp\langle 12|V|2'1'\rangle=
\langle 12|V|1'2'\rangle
\mp\langle 21|V|1'2'\rangle
\end{equation}
with the upper sign for fermions and the lower sign for bosons.  
Insertion of Eq.~(\ref{4.1}) into Eqs.~(\ref{2.13}) and~(\ref{2.14}) 
gives
\begin{eqnarray}
& &
\label{2.x1}
\varrho^{}_{\rm rel}(t)={1\over Z^{}_{\rm rel}(t)}\,
\exp\left\{-\beta^*(t)\,\hat{\cal H'}(t) 
- \sum_{1} \widetilde\Lambda^{}_1(t)\,\hat f^{}_1 \right\},
\\[6pt]
& &
\label{2.x2}
Z^{}_{\rm rel}(t)={\rm Tr}\,
\exp\left\{-\beta^*(t)\,\hat{\cal H}'(t)  
- \sum_{1} \widetilde\Lambda^{}_1(t)\,\hat f^{}_1 \right\}.
\end{eqnarray}
The new Lagrange multipliers, $\widetilde\Lambda^{}_1(t)$,
are related to $\Lambda^{}_1(t)$ and $\lambda^{}_1(t)$ by
\begin{equation}
\label{2.x3}
\widetilde\Lambda^{}_1(t)=\Lambda^{}_1(t)+\beta^*(t)\,
\Sigma^{\rm HF}_{1}(t)=\beta^*(t)\,\left[E^{}_1(t)
-\mu^*(t)\right] +\lambda^{}_1(t). 
\end{equation}
We emphasize once again that the representations for 
the relevant statistical operator given by 
Eqs.~(\ref{2.13}),~(\ref{2.9}), and~(\ref{2.x1}),
are equivalent to each other and differ only in 
the definition of the
Lagrange multipliers conjugated to the one-particle 
distribution function.  

\subsection{Thermodynamic relations}
It is important to note that the relevant statistical operator
allows to extend thermodynamic relations to
nonequilibrium systems (see, e.g.,~\cite{ZubMorRoep1}).    
The key quantities are the Massieu-Planck function
\begin{equation}
\label{2.16}
\Phi(t)= \ln Z^{}_{\rm rel}(t)= \ln {\rm Tr}\,
\exp\left\{
-\sum_{m} \lambda^{}_m(t)\,\hat P^{}_m
\right\}
\end{equation}
and the nonequilibrium entropy 
\begin{equation}
\label{2.x4}
S(t)=-{\rm Tr}
\left\{\varrho^{}_{\rm rel}(t)\,\ln\varrho^{}_{\rm rel}(t)\right\}
=\Phi(t) +\sum^{}_m \lambda^{}_m(t)\,P^{}_m(t),
\end{equation}
which play the role of  thermodynamic potentials in the variables
$\{\lambda^{}_m(t)\}$ and $\{ P^{}_m(t)\}$, 
respectively.
In the case under consideration, using the above given  
representations for the partition function, we obtain
formally different but equivalent thermodynamic relations. 
Taking, for instance, 
the partition function in the form~(\ref{2.14}), we see that the
Massieu-Planck function can be interpreted as
a nonequilibrium thermodynamic potential in the 
variables $\beta^*(t)$,
$\mu^*(t)$, and $\lambda^{}_1(t)$. 
Calculation of its variation gives
\begin{equation}
\label{2.18}
\delta \Phi(t)=
-\big[{\cal E}-\mu^*(t)N\big]\,\delta\beta^*(t)
+\beta^*(t) N\, \delta \mu^*(t)
- \sum^{}_1 f^{}_1(t)\,\delta\lambda^{}_1(t). 
\end{equation}
In the same representation, 
the entropy~(\ref{2.x4}) can be written as
\begin{equation}
\label{2.17}
S(t)=\Phi(t) + \beta^*(t)\,\big[{\cal E} -\mu^*(t)N\big]
+\sum_{1} \lambda^{}_1(t) f^{}_1(t),
\end{equation}
where  the self-consistency conditions~(\ref{2.15}) have been used.
Varying both sides of Eq.~(\ref{2.17}) and recalling Eq.~(\ref{2.18}),
we find
\begin{eqnarray}
\label{2.19}
\delta S(t)&=& \beta^*(t)\, \big[\delta{\cal E} - \mu^*(t)\,\delta N\big]
+\sum_{1}\lambda^{}_1(t)\,\delta f^{}_1(t)
\nonumber\\[10pt]
{}&=& 
\beta^*(t)\,\delta{\cal E} +
\sum_{1}\left[\lambda^{}_1(t) 
-\beta^*(t)\,\mu^*(t)\right]\delta f^{}_1(t).
\end{eqnarray}
In the last line we have taken into account that 
$\delta N=\sum_{1} \delta f^{}_1$. The entropy
may thus be considered as a nonequilibrium thermodynamic 
potential in the  variables ${\cal E}$ and $f^{}_1(t)$ or,
what is the same --- in the variables 
${\cal E}^{}_{\rm int}(t)$ and $f^{}_1(t)$. 
>From  Eqs.~(\ref{2.19}) it follows that
\begin{equation}
\label{2.20}
\beta^*(t)=
\left(\frac{\partial S(t)}{\partial{\cal E}}\right)^{}_{\!\!f}
= \left(\frac{\partial S(t)}
{\partial{\cal E}^{}_{\rm int}(t)}\right)^{}_{\!\!f},
\qquad
\lambda^{}_1(t)-\beta^*(t)\,\mu^*(t)=
\left(\frac{\partial S(t)}
{\partial f^{}_1(t)}\right)^{}_{\!\!{\cal E}},
\end{equation} 
The second equation confirms the fact 
that thermodynamics 
determines only the combination of the Lagrange multiplier
$\lambda^{}_1(t)$ and the quasi-chemical potential.

\subsection{Evolution equations for thermodynamic quantities}
As already discussed, in the approach presented here 
the total energy is regarded as an independent state parameter
in addition to the one-particle function . 
The evolution equation for the total energy is trivial: 
$d{\cal  E}/dt=0$. We shall see, however, that  the correlation
contribution to the kinetic equation  is related to the time
dependence of other thermodynamic quantities, such as the 
quasi-temperature and the interaction energy.
We will now show that the evolution equations for the thermodynamic
quantities of interest can be expressed in terms of the
collision integral~(\ref{2.6}).

\subsubsection{Energy balance}
We start with equation for the interaction 
energy, Eq.~(\ref{2.8}).  Since the total energy is conserved and
\begin{equation}
\label{3.1}
{\cal E}=\langle \hat H\rangle^t=\sum_{1}\varepsilon^{}_1\,f^{}_1(t)  
+{\cal E}^{}_{\rm int}(t),
\end{equation}
we immediately obtain the balance equation
\begin{equation}
\label{3.2}
\frac{d{\cal E}^{}_{\rm int}(t)}{dt}=
-\sum_{1} \varepsilon^{}_1\,I^{}_1(t).
\end{equation}
In dealing with nonequilibrium  many-particle correlations,
it  is  convenient to introduce the  {\em correlation energy},
${\cal E}^{}_{\rm corr}(t)$,  which is  defined as
\begin{equation}
\label{3.3}
{\cal E}^{}_{\rm corr}(t)= 
{\cal E}^{}_{\rm int}(t) - {\cal E}^{}_{\rm HF}(t)=
{\cal E} - \sum_{1}\varepsilon^{}_1\,f^{}_1(t)
- {\cal E}^{}_{\rm HF}(t),
\end{equation} 
where
\begin{equation}
\label{3.4}
{\cal E}^{}_{\rm HF}(t)=
{1\over2}\sum_{12} \langle 12|V|12\rangle^{}_{\rm ex}\,
f^{}_1(t) f^{}_2(t) =
{1\over 2} \sum^{}_1 \Sigma^{\rm HF}_1(t)\,f^{}_{1}(t)
\end{equation}
is the Hartree-Fock contribution to the total energy.
>From Eqs.~(\ref{3.2}) and (\ref{3.3}) follows the balance equation 
\begin{equation}
\label{3.6}
\frac{d{\cal E}^{}_{\rm corr}(t)}{dt}=
-\sum_{1} E^{}_1(t)\,I^{}_1(t)
\end{equation}
which differs from Eq.~(\ref{3.2}) in that now the 
single-particle are given by Eq.~(\ref{3.7}),
i.e., they involve  the  exchange Hartree-Fock term.
 
\subsubsection{Equation for the quasi-temperature}
In principle, the evolution equation for $\beta^*(t)$
may be derived from the equation of state, 
$\beta^*(t)=\beta^*\left( {\cal E}, \{f(t)\} \right)$, where
the second argument indicates  that $\beta^*$  is a functional of 
$f^{}_1(t)$. 
This way, however, is not appropriate because we have to 
calculate the 
functional derivative of $\beta^*(t)$ with respect to the
one-particle distribution function.
It is more convenient to make use of the self-consistency 
conditions~(\ref{2.11}) by differentiating them with respect to
time. This gives  
\begin{equation}
\label{3.9}
{\rm Tr}\left\{\frac{\partial\varrho^{}_{\rm rel}(t)}{\partial t}\, 
\hat f^{}_1\right\}=I^{}_1(t),
\qquad
{\rm Tr}\left\{\frac{\partial\varrho^{}_{\rm rel}(t)}{\partial t}\, 
\hat H'\right\}=\frac{d{\cal E}^{}_{\rm int}(t)}{dt}.
\end{equation}
Since the relevant statistical operator, 
when taken in the form~(\ref{2.9}), 
depends on time through the Lagrange multipliers $\beta^*(t)$ and
$\Lambda^{}_1(t)$, we may write
 \begin{equation}
  \label{3.10}
\frac{\partial\varrho^{}_{\rm rel}(t)}{\partial t}=
 \frac{\delta\varrho^{}_{\rm rel}(t)}{\delta\beta^*(t)}\,
\frac{d\beta^*(t)}{dt}+
\sum_1 \frac{\delta\varrho^{}_{\rm rel}(t)}{\delta\Lambda^{}_1(t)}\,
\frac{\partial\Lambda^{}_1(t)}{\partial t}.  
\end{equation}
With the aid of Eqs.~(\ref{2.9}) and~(\ref{2.10}), the 
variations of $\varrho^{}_{\rm rel}(t)$ can easily be  calculated
and then Eqs.~(\ref{3.9}) are  transformed to  
(for brevity, the argument $t$ is omitted)
\begin{eqnarray}
  \label{3.11}
& &
\left(\hat f^{}_1,\hat H'\right) \frac{d\beta^*}{dt}
+\sum_{1'} \left(\hat f^{}_1,\hat f^{}_{1'}\right)
\frac{\partial\Lambda^{}_{1'}}{\partial t}=-I^{}_1,
\\[10pt]
\label{3.12}
& &
\left(\hat H',\hat H'\right) \frac{d\beta^*}{dt}
+\sum_{1} \left(\hat H',\hat f^{}_1 \right) 
\frac{\partial\Lambda^{}_{1}}{\partial t}=
-\frac{d{\cal E}^{}_{\rm int}}{dt}, 
\end{eqnarray}
where we have introduced the correlation function of two
dynamical variables:
\begin{equation}
  \label{3.13}
\left(\hat A,\hat B \right)=
\int_0^1 dx\,\left\langle \Delta \hat A\,\varrho^{x}_{\rm rel}\,
\Delta\hat B\,\varrho^{-x}_{\rm rel}\right\rangle^{}_{\rm rel}   
\end{equation}
with $\Delta \hat A=\hat A -\langle\hat A\rangle^{}_{\rm rel}$.
Equation~(\ref{3.11}) can formally be solved for 
$\partial\Lambda^{}_1/\partial t$ to give
\begin{equation}
  \label{3.14}
\frac{\partial\Lambda^{}_1}{\partial t}=
- \sum_{1'} \chi^{-1}_{11'}
\left\{
\left(\hat f^{}_{1'},\hat H'\right)\frac{d\beta^*}{dt}
+I^{}_{1'}
\right\},  
\end{equation}
where $\chi^{-1}$ is inverse to the correlation matrix
\begin{equation}
\label{3.15}
\chi^{}_{11'}= \left(\hat f^{}_1,\hat f^{}_{1'}\right). 
\end{equation}
Substituting the expression~(\ref{3.14}) into Eq.~(\ref{3.12})
and recalling the balance equation~(\ref{3.2}),
we obtain
\begin{equation}
\label{3.16}
\frac{d\beta^*(t)}{dt}=
\frac{1}{C}\sum_{11'} \left(\hat H,\hat f^{}_1\right)
\chi^{-1}_{11'}\,I^{}_{1'}
\end{equation}
with the following notation:
\begin{equation}
\label{3.17}
C=
\left(\hat H',\hat H'\right)
-\sum_{11'}\left(\hat H',\hat f^{}_1\right)
\chi^{-1}_{11'}\left(\hat f^{}_{1'},\hat H'\right).     
\end{equation}
It is easy to check that $C$ in invariant under transformations 
$\hat H'\to \hat H'+\sum_{1} \alpha^{}_1 \hat f^{}_1$
with arbitrary coefficients
$\alpha^{}_k$. This property allows one,
for instance, to replace $\hat H'$ in  Eq.~(\ref{3.17}) by the 
operator
$\hat{\cal H}'(t)$ [cf. Eqs.~(\ref{4.2})].

\subsubsection{The entropy production}
We finally derive the entropy balance equation. Recalling
Eq.~(\ref{2.19}), we write
$$
\frac{dS(t)}{dt}=\beta^*(t)\,\frac{d{\cal E}(t)}{dt}
+ \sum_{1} \left[\lambda^{}_1(t) -\beta^*(t)\,\mu^*(t)\right]
\frac{\partial f^{}_1(t)}{\partial t}.
$$
The first term on the right-hand side is zero since the total
energy is conserved. Taking also into account that 
$\partial f^{}_1/\partial t=I^{}_1$ and 
$\sum_{1} I^{}_1=dN/dt=0$, we arrive at the equation
\begin{equation}
\label{3.19}
\frac{dS(t)}{dt}= \sum_{1} \lambda^{}_1(t)\,I^{}_1(t)
\end{equation} 
which determines the entropy production
in the system. We have already noted that, in thermal equilibrium,
the Lagrange multipliers $\lambda^{}_1(t)$ are equal to zero.
In addition, the collision integral is  also zero in thermal
equilibrium. Consequently, the entropy production given by 
Eq.~(\ref{3.19}) is at least of second order in the deviations 
from equilibrium, as it should be.
 
\setcounter{equation}{0}

\section{The Nonequilibrium statistical operator}
To proceed beyond the formal thermodynamic relations,
we have to calculate the collision 
integral~(\ref{2.6}). In other words, we have to
find a solution of the von Neumann equation~(\ref{2.4})
for the statistical operator $\varrho(t)$ in terms of the state
parameters.  
Since Eq.~(\ref{2.4}) is a differential equation with 
respect to time, 
one has to give the statistical operator at some initial 
time $t^{}_0$
or require the $\varrho(t)$ to satisfy some boundary condition  
for $t\to -\infty$. An appropriate choice of the  
initial or boundary condition depends on the physical situation
under consideration. 
Formally, one may consider 
the initial statistical operator $\varrho(t_0)$ 
or the limiting statistical operator in a distant past 
to be arbitrary. We will, however, 
use the initial condition 
$\varrho(t^{}_0)=\varrho^{}_{\rm rel}(t^{}_0)$ at some 
time $t^{}_0$
and the boundary condition that the true statistical operator 
coincides with the relevant
statistical operator in a distant past. 
The choice of these special conditions, however, is 
taken only for simplicity since it is
of little consequence for
the long-time behavior of the kinetic equation, which    
is of interest to us here.

Assuming that
\begin{equation}
\label{4.4}
\varrho(t^{}_0)=\varrho^{}_{\rm rel}(t^{}_0)
\end{equation}
and using the decomposition~(\ref{4.1}) of the Hamiltonian,
it can easily be verified that the von Neumann 
equation~(\ref{2.4}) is equivalent to the integral equation
\begin{eqnarray}
\label{4.5}
& &
\hspace*{-40pt}
\varrho(t)=\varrho^{}_{\rm rel}(t)
{}- \int_{t_0}^t dt'\,
U^{}_0(t,t')
\left\{
\frac{\partial\varrho^{}_{\rm rel}(t')}{\partial t'}
+
{1\over i\hbar}\left[\varrho^{}_{\rm rel}(t'), \hat{\cal H}_0(t')\right]
\right\} U^{\dagger}_0(t,t')
\nonumber\\[8pt]
& &
\hspace*{90pt}
{}- \int_{t_0}^t dt'\,
U^{}_0(t,t')\,
{1\over i\hbar}\left[\varrho(t'), \hat{\cal H'}(t')\right]
U^{\dagger}_0(t,t'),
\end{eqnarray}
where
\begin{equation}
\label{4.6}
U^{}_0(t,t')= \exp^{}_+\left\{-{i\over\hbar}\int_{t'}^t     
\hat{\cal H}^{}_0(t'')\,dt''
\right\}
\end{equation}
is the unperturbed evolution operator;
the symbol $\exp^{}_{+}\left\{\cdots\right\}$ stands for the
time-ordered exponent. Equation~(\ref{4.5}) is still exact.
If $\hat{\cal H}'(t)$ is treated as a small perturbation, an
approximate solution of Eq.~(\ref{4.5})  
can be found by an iterative procedure. 
We shall restrict our discussion to the second-order  
non-Markovian Born approximation in the collision integral,
which leads, in the standard approach~\cite{HaugJauho96},
to the Levinson kinetic equation.

To calculate the second-order collision
integral~(\ref{2.6}), we need the statistical 
operator $\varrho(t)$,
correct to first order in the perturbation $\hat{\cal H}'$.  
We shall show  later than the time derivative 
$\partial\varrho^{}_{\rm rel}(t')/\partial t'$ 
in Eq.~(\ref{4.5}) is at least of second
order in $\hat{\cal H}'$ and, consequently, can be omitted.
We next note that the interaction term enters 
explicitly into the relevant 
statistical operator~[see, e.g.,~(\ref{2.x1})]. 
Therefore the leading interaction  contribution to the
commutator 
$[\varrho^{}_{\rm rel}(t'),\hat{\cal H}^{}_0(t')]$ is 
at least linear in  $\hat{\cal H}'$; this term must be
retained. 
Finally, since the last term in Eq.~(\ref{4.5}) is already of
first order in the interaction, we  
may replace $\varrho(t')$ in this
term by $\varrho^{}_{\rm rel}(t')$.
Thus, the nonequilibrium statistical operator, correct to
first order in $\hat{\cal H}'$, is given by
\begin{eqnarray}
\label{4.7}
\varrho(t)=\varrho^{}_{\rm rel}(t)
& &{}- \int_{t_0}^t dt'\,
U^{}_0(t,t')\,
{1\over i\hbar}\left[\varrho^{}_{\rm rel}(t'),\hat{\cal H}^{}_0(t')\right]
\,U^{\dagger}_0(t,t')
\nonumber\\[8pt]
& &{}- \int_{t_0}^t dt'\,
U^{}_0(t,t')\,
{1\over i\hbar}\left[\varrho^{}_{\rm rel}(t'),\hat{\cal H}'(t')\right]
\,U^{\dagger}_0(t,t').
\end{eqnarray}

Another solution of the von Neumann equation can be obtained by
imposing the boundary condition that the true nonequilibrium
statistical operator coincides with the relevant 
statistical operator
in a distant past. 
This solution follows easily by adding
to Eq.~(\ref{2.4}) an infinitesimally small 
source~\cite{ZubMorRoep1}:
\begin{equation}  
\label{4.8}
\frac{\partial\varrho(t)}{\partial t}+
{1\over i\hbar}\left[\varrho(t),\hat{\cal H}^{}_0(t)+
\hat{\cal H}'(t)
\right]= -\varepsilon
\left\{\varrho(t)-\varrho^{}_{\rm rel}(t)\right\},  
\end{equation}
where $\varepsilon\to +0$ after the calculation of averages with 
$\varrho(t)$. 
It is important to note that the inclusion of the source 
term into the von Neumann equation does not violate the energy 
conservation. 
This can be seen by taking the product of Eq.~(\ref{4.8})
with $\hat H$ and calculating the trace. Then we obtain
\begin{equation}
\label{4.8x}
\frac{d{\cal E}(t)}{dt}= - \varepsilon\left\{   
{\cal E}(t) - {\rm Tr}\left( \hat H\varrho^{}_{\rm rel}(t)\right)
\right\}.
\end{equation}
The right-hand side of this equation is zero due to the
self-consistency condition for the total energy in the 
relevant ensemble.

Analogously to  the derivation of Eq.~(\ref{4.7}), we use
Eq.~(\ref{4.8}) to derive the  
first-order statistical operator in the form 
\begin{eqnarray}
\label{4.9}
\varrho(t)=\varrho^{}_{\rm rel}(t)
& &{}- \int_{-\infty}^t dt'\, {\rm e}^{-\varepsilon(t-t')}\,
U^{}_0(t,t')\,
{1\over i\hbar}
\left[\varrho^{}_{\rm rel}(t'),\hat{\cal H}^{}_0(t')\right]
\,U^{\dagger}_0(t,t')
\nonumber\\[8pt]
& &{}- \int_{-\infty}^t dt'\, {\rm e}^{-\varepsilon(t-t')}\,
U^{}_0(t,t')\,
{1\over i\hbar}
\left[\varrho^{}_{\rm rel}(t'),\hat{\cal H}'(t')\right]
\,U^{\dagger}_0(t,t').
\end{eqnarray}
This expression can also be interpreted as a 
rule for passing to the limit
$t^{}_0\to-\infty$ in Eq.~(\ref{4.7}) since the factor 
$\exp\{-\varepsilon(t-t')\}$ provides the regularization of the 
integral.

\setcounter{equation}{0}

\section{The non-Markovian collision integral}
We now turn to the calculation of the collision integral~(\ref{2.6})
using the explicit expression~(\ref{4.7}) for the 
statistical operator.
First we will show that the term 
$\varrho^{}_{\rm rel}(t)$ in Eq.~(\ref{4.7})
does not contribute to the collision integral.
Note that  the obvious identity
$\big\langle \big[\hat f^{}_1,\ln \varrho^{}_{\rm rel}(t)  
\big]\big\rangle^t_{\rm rel}=0$ and  
Eq.~(\ref{2.9}) give
$$
\beta^*(t)
\big\langle \big[\hat f^{}_1,\hat{H}'\big]\big\rangle^t_{\rm rel}
+\sum_{1'} \Lambda^{}_{1'}(t)
\big\langle \big[\hat f^{}_{1},\hat f^{}_{1'}  
\big]\big\rangle^t_{\rm rel}=0.
$$
Since 
$ \big[\hat f^{}_{1},\hat f^{}_{1'}\big]=0$, we find that
$\big\langle 
\big[\hat f^{}_1,\hat H'\big]\big\rangle^t_{\rm rel}=0$.   
Thus the collision
integral~(\ref{2.6}) is at least of second order in the 
interaction.
This allows us to show that the time derivative 
$\partial\varrho(t')/\partial t'$ in Eq.~(\ref{4.5}) is 
also of second order
in $\hat{\cal H}'$ and, consequently, it does not contribute to the
first-order expressions~(\ref{4.7}) and~(\ref{4.9}).
We use the fact  that the relevant statistical operator depends
on time through the state parameters or through the conjugated
Lagrange multipliers.  
For instance, we may assume that
$\varrho^{}_{\rm rel}(t')=\varrho^{}_{\rm rel}({\cal E},\{f(t')\})$.
Then, since the total energy is conserved, 
$$
\frac{\partial\varrho^{}_{\rm rel}(t')}{\partial t'}=
\sum_{1} \frac{\delta\varrho^{}_{\rm rel}(t')}{\delta f^{}_1(t')}\,
\frac{\partial f^{}_1(t')}{\partial t'}=
\sum_{1} \frac{\delta\varrho^{}_{\rm rel}(t')}{\delta f^{}_1(t')}\,
I^{}_1(t'),
$$
whence it follows that 
$\partial\varrho^{}_{\rm rel}(t')/\partial t'$
is at least of second order in the perturbation,
as was to be proved.

The last two terms in the expression~(\ref{4.7}), 
when substituted into
Eq.~(\ref{2.6}), lead to the  
decomposition of the collision
integral
\begin{equation}
\label{5.1}
I^{}_1(t)=I^{L}_{1}(t) + I^{C}_{1}(t),
\end{equation}
where
\begin{eqnarray}
\label{5.2}
& &
\hspace*{-30pt}
I^{L}_1(t)=-{1\over\hbar^2}
\int_{t_0}^{t} dt'\, 
{\rm Tr}\left\{
\big[ U^{\dagger}_0(t,t')\big[ \hat f^{}_1,\hat H'\big]
U^{}_0(t,t'),\hat{\cal H}'(t')\big]
\varrho^{}_{\rm rel}(t')
\right\},
\\[8pt]
\label{5.3}
& &
\hspace*{-30pt}
I^{C}_1(t)={1\over\hbar^2}
\int_{t_0}^{t} dt'\, 
{\rm Tr}
\left\{
 U^{\dagger}_0(t,t') \big[\hat f^{}_1,\hat H'\big]
U^{}_0(t,t') 
\big[\varrho^{}_{\rm rel}(t'),\hat{\cal H}^{}_0(t')\big]
\right\}.
\end{eqnarray} 
We shall see later that the term $I^L_1(t)$
is nothing but the Levinson collision integral. 
The new term, $I^C_1(t)$, is due to many-particle 
correlations in the 
ensemble described by the relevant statistical operator.  
If the one-particle distribution function 
$f^{}_1(t)$ is taken as the only state parameter, the relevant
statistical operator~(\ref{2.9}) does not involve the term 
with $\hat H'$; hence, $\varrho^{}_{\rm rel}(t')$ commutes with 
$\hat{\cal H}^{}_0(t')$ and $I^C_1(t)=0$. 

\subsection{The collision contribution}
We now calculate the term $I^L_1(t)$, Eq.~(\ref{5.2}), in the
non-Markovian Born approximation. 
Since  this term  is explicitly of second order in the 
interaction, the relevant statistical
operator, Eq.~(\ref{2.x1}), can be approximated by
\begin{equation}
  \label{5.4}
  \varrho^0_{\rm rel}(t)=
\exp\left\{
-\sum_1 \widetilde\Lambda^{}_1(t)\,a^{\dagger}_1 a^{}_1\right\}
\left/
{\rm Tr}\,
\exp\left\{-\sum_1 \widetilde\Lambda^{}_1(t)\,
a^{\dagger}_1 a^{}_1\right\}.
\right.
\end{equation}
Formally, this statistical operator describes a 
nonequilibrium ideal quantum gas, so that the self-consistency
condition for the Lagrange multipliers 
$\widetilde\Lambda^{}_1(t)$ reads
\begin{equation}
  \label{5.6}
f^{}_1(t)=
{\rm Tr}\left\{a^{\dagger}_1 a^{}_1
\varrho^{0}_{\rm rel}(t) \right\}=
\frac{1}{\exp\left\{\widetilde\Lambda^{}_1(t)\right\} \pm 1},
\end{equation}   
whence it follows that
\begin{equation}
\label{5.19x}
\widetilde\Lambda^{}_1(t)=
\ln\left(\frac{1\mp f^{}_1(t)}{f^{}_1(t)}\right).
\end{equation}
The time dependence of the operators in Eq.~(\ref{5.2}) can be worked
out by using the following properties of the evolution 
operator~(\ref{4.6}):
\begin{equation}
\label{5.9}
U^{\dagger}_0(t,t')\,a^{}_1\,U^{}_0(t,t')=
{\rm e}^{-i\omega^{}_1(t,t')}\,a^{}_1,
\qquad
U^{\dagger}_0(t,t')\,a^{\dagger}_1\,U^{}_0(t,t')=
{\rm e}^{i\omega^{}_1(t,t')}\,a^{\dagger}_1,
\end{equation}
where
\begin{equation}
\label{5.10}
\omega^{}_1(t,t')={1\over\hbar}\int_{t'}^{t} dt''\,E^{}_1(t'')=
{\varepsilon^{}_1\over\hbar}(t-t') + {1\over\hbar}\int_{t'}^{t} dt''\,
\Sigma^{{\rm HF}}_1(t'').
\end{equation}
Taking into account that the statistical operator~(\ref{5.4})
admits Wick's decomposition of the averages,
a simple algebra gives 
\begin{equation}
  \label{5.11}
I^L_1(t)=
-{1\over\hbar^2}
\sum_{21'2'} \left|\langle 12|V|1'2'\rangle^{}_{\rm ex}\right|^2
\int_{t_0}^{t} dt'\,
\cos\left[\Delta \omega^{}_{12,1'2'}(t,t')\right]  
{\cal F}^{}_{12,1'2'}\left(\{f(t')\}\right),
\end{equation}
where
\begin{equation}
\label{5.12}
\Delta \omega^{}_{12,1'2'}(t,t')=\omega^{}_{1}(t,t')
+\omega^{}_{2}(t,t')-\omega^{}_{1'}(t,t')-\omega^{}_{2'}(t,t'),
\end{equation}
and the functional ${\cal F}^{}_{12,1'2'}\left(\{\varphi\}\right)$ 
is defined for any set of single-particle functions  
$\varphi^{}_1$ as
\begin{equation}
\label{5.13}
{\cal F}^{}_{12,1'2'}\left(\{\varphi\}\right)=
\varphi^{}_1\,\varphi^{}_2\,\bar\varphi^{}_{1'}\, \bar\varphi^{}_{2'}
- \bar\varphi^{}_{1}\,\bar\varphi^{}_{2}\,\varphi^{}_{1'}\,
\varphi^{}_{2'},
\qquad
\bar\varphi^{}_1=1 \pm  \varphi^{}_1.
\end{equation}
In the context of Eq.~(\ref{5.11}), the functional 
${\cal F}^{}_{12,1'2'}\left(\{f(t')\}\right)$ is nothing but the 
gain-loss term which appears in quantum collision 
integrals. 
Expression~(\ref{5.11}) only differs from the original
Levinson collision integral~\cite{Levinson69} in that 
the quantities $\omega^{}_k(t,t')$ in 
Eq.~(\ref{5.12}) involve the exchange Hartree-Fock term.

\subsection{The correlation contribution}
To calculate the  second-order correlation contribution to
the collision integral, Eq.~(\ref{5.3}), 
we  expand the relevant statistical 
operator~(\ref{2.x1}) in $\hat{\cal H}'$,
keeping  only the first-order terms. This gives 
\begin{equation}
\label{5.14}
\varrho^{}_{\rm rel}(t)=
\left\{
1 - \beta^*(t)
\int_0^1 dx \left[\varrho^0_{\rm rel}(t)\right]^x
\left(
\hat{\cal H}'(t) -\langle \hat{\cal H}'\rangle^{t}_{0}
\right)
\left[\varrho^0_{\rm rel}(t)\right]^{-x}
\right\}  
\varrho^0_{\rm rel}(t).
\end{equation}
Here  the symbol $\langle\hat{\cal H}'\rangle^{t}_{0}$
stands for the average  with the statistical 
operator~(\ref{5.4}). Having  the above expression, we calculate
the commutator appearing in Eq.~(\ref{5.3}) 
(for brevity the time argument $t'$ is omitted):
\begin{equation}
  \label{5.15}
\left[\varrho^{}_{\rm rel},\hat{\cal H}^{}_0\right]=
-{1\over2}\,\beta^* \sum_{121'2'}
\langle 1'2'|V|12\rangle \,
\frac{\Delta E^{}_{12,1'2'}}{\Delta\widetilde\Lambda^{}_{12,1'2'}}
\left\{{\rm e}^{\Delta\widetilde\Lambda^{}_{12,1'2'}}-1\right\}
a^{\dagger}_{2'}a^{\dagger}_{1'} a^{}_{1} a^{}_2\,
\varrho^0_{\rm rel}, 
\end{equation}
where
\begin{eqnarray}
& &
\label{5.16}
\Delta E^{}_{12,1'2'}(t)= E^{}_1(t) + E^{}_1(t) 
- E^{}_{1'}(t) - E^{}_{2'}(t),
\\[6pt]
\label{5.17}
& &
\Delta\widetilde\Lambda^{}_{12,1'2'}(t)=
\widetilde\Lambda^{}_{1}(t)+  
\widetilde\Lambda^{}_{2}(t)-
\widetilde\Lambda^{}_{1'}(t)-
\widetilde\Lambda^{}_{2'}(t). 
\end{eqnarray}
We next substitute Eq.~(\ref{5.15}) into Eq.~(\ref{5.3})
and  use  Wick's theorem to calculate the average. 
In the final result it is convenient
to eliminate $\Delta\widetilde\Lambda^{}_{12,1'2'}$. 
 To this end, we introduce  a functional
\begin{equation}
  \label{5.18}
 {\cal K}^{}_{12,1'2'}\left(\{\varphi\}\right)=
 \frac{\varphi^{}_1\,\varphi^{}_2\,
\bar\varphi^{}_{1'}\,\bar\varphi^{}_{2'}}
{\bar\varphi^{}_1\,\bar\varphi^{}_2\,\varphi^{}_{1'}\,\varphi^{}_{2'}}.
\end{equation}
Then, recalling Eq.~(\ref{5.19x}), it can easily be verified that
\begin{equation}
  \label{5.19}
\Delta\widetilde\Lambda^{}_{12,1'2'}(t)=
-\ln{\cal K}^{}_{12,1'2'}\left(\{f(t)\} \right).
\end{equation}
Omitting a simple algebra, we present the final expression for the
correlation term in the collision integral:
\begin{eqnarray}
  \label{5.21}
& &
\hspace*{-10pt}
I^C_1(t)=
- {1\over\hbar^2}
\sum_{21'2'} \big|\langle 12|V|1'2'\rangle^{}_{\rm ex}\big|^2
\int_{t_0}^{t} dt'\,
\cos\left[\Delta\omega^{}_{12,1'2'}(t,t'\right]
\nonumber\\[8pt]
& &
\hspace*{160pt}
{}\times
\frac{\beta^*(t')\,\Delta E^{}_{12,1'2'}(t')}
{\ln{\cal K}^{}_{12,1'2'}\left(\{f(t')\} \right)}\,  
{\cal F}^{}_{12,1'2'}\left(\{f(t')\}\right).
\end{eqnarray}
It is similar to the collision  term~(\ref{5.11}) but
contains the additional factor in the integrand.

\subsection{The full collision integral and its properties}
Due to the similarity in structure, the two contributions, 
Eqs.~(\ref{5.11}) and~(\ref{5.21}),
are conveniently combined into a single expression 
\begin{eqnarray}
  \label{5.22x}
& &
I^{}_1(t)= -
{1\over\hbar^2}
\sum_{21'2'} \left|\langle 12|V|1'2'\rangle^{}_{\rm ex}\right|^2
\int_{t_0}^{t} dt'\,
\cos\left[\Delta\omega^{}_{12,1'2'}(t,t')\right]
\nonumber\\[9pt]
& &
\hspace*{160pt}
{}\times
\left\{
1+
\frac{\beta^*(t')\,\Delta E^{}_{12,1'2'}(t')}
{\ln{\cal K}^{}_{12,1'2'}\left(\{f(t')\} \right)}
\right\}  
{\cal F}^{}_{12,1'2'}\left(\{f(t')\}\right),
\end{eqnarray} 
which can be written in a more
elegant form by using the relation
\begin{equation}
\label{5.20}
\beta^*(t)\,\Delta E^{}_{12,1'2'}(t)=
-\ln{\cal K}^{}_{12,1'2'}\left(\{F(t)\} \right), 
\end{equation}
where
\begin{equation}
  \label{5.8}
F^{}_1(t)=
\frac{1}{\exp\left\{\beta^*(t)\left[E^{}_1(t)
-\mu^*(t)\right]\right\}\pm 1}  
\end{equation}
is the one-particle distribution function
in the quasi-equilibrium
ensemble~[cf. Eq.~(\ref{2.22})]. With Eq.~(\ref{5.20}),
the expression~(\ref{5.22x}) becomes
\begin{eqnarray}
  \label{5.22}
& &
I^{}_1(t)= -
{1\over\hbar^2}
\sum_{21'2'} \left|\langle 12|V|1'2'\rangle^{}_{\rm ex}\right|^2
\int_{t_0}^{t} dt'\,
\cos\left[\Delta\omega^{}_{12,1'2'}(t,t')\right]
\nonumber\\[9pt]
& &
\hspace*{160pt}
{}\times
\left\{
1-
\frac{\ln{\cal K}^{}_{12,1'2'}\left(\{F(t')\} \right)}
{\ln{\cal K}^{}_{12,1'2'}\left(\{f(t')\} \right)}
\right\}  
{\cal F}^{}_{12,1'2'}\left(\{f(t')\}\right).
\end{eqnarray} 
This collision integral  has some remarkable properties. First, 
it vanishes in the quasi-equilibrium state where $f^{}_1(t)=F^{}_1(t)$.
In this case the collision contribution 
and the correlation contribution cancel each other.
In particular, $I^{}_1=0$ in complete equilibrium since
$F^{}_1(t)$ goes over to the equilibrium distribution function as 
$\beta^*(t)\to 1/T$ and $\mu^*(t)\to\mu$, 
where $T$ and $\mu$ are
the equilibrium temperature and the equilibrium
chemical potential, respectively.
It should be emphasized  that the collision term~(\ref{5.11})
alone does not vanish in thermal equilibrium, which is the grave
disadvantage of the Levinson-type kinetic equations. 
Another important property of the collision integral 
is its asymptotic behavior as $t -t^{}_0\to\infty$.
This stage of the evolution can be described on 
a large time scale, so that we may pass  to the Markovian 
limit. To analyze this case, it is convenient to return to
the expression~(\ref{5.22x}).
Putting  there $f(t')\approx f(t)$, $F(t')\approx F(t)$, 
and then performing the limit $t^{}_0\to -\infty$ with the 
regularization factor $\exp\left\{ -\varepsilon(t-t')\right\}$,
we find that the correlation contribution
vanishes due to the fact that now the cosine 
term is replaced by 
the delta function 
$\delta\left(\Delta E^{}_{12,1'2'}(t)/\hbar\right)$.
As a result, we get the well-known Uehling-Uhlenbeck 
collision integral 
\begin{equation}
  \label{5.24}
I^{}_1(t)= -
{\pi\over\hbar^2}
\sum_{21'2'} \left|\langle 12|V|1'2'\rangle^{}_{\rm ex}\right|^2\,
\delta\left(\frac{\Delta E^{}_{12,1'2'}(t)}{\hbar}\right)
\left(f^{}_1 f^{}_2 \bar f^{}_{1'}\bar f^{}_{2'}
- \bar f^{}_{1}\bar f^{}_{2}f^{}_{1'} f^{}_{2'}
 \right)^{}_{t},  
\end{equation}
where we have used the definition of 
${\cal F}^{}_{12,1'2'}$, Eq.~(\ref{5.13}). 
We would like to  draw
attention to the role of the correlation term in the collision
integral. 
Although this term goes to zero in the long-time limit, the
non-Markovian expression~(\ref{5.22x}) is constructed such that 
the interplay between collisions and correlations is precisely
the reason why the Markovian regime arises.
It should be noted, however, that beyond Born approximation, for
instance, in the $T$-approximation for the collision integral,
the correlation term does not go to zero in the Markovian 
limit~\cite{MorozovRoepke95}. 
 
\subsection{A simplified version of the non-Markovian 
collision integral}
Because of the presence of the ${\cal K}$-functional, the full
collision integral~(\ref{5.22x})
has a more complicated structure than the Levinson 
term~(\ref{5.11}).  
Having in mind practical applications of the scheme developed
here, it makes sense to formulate a simplified version of the 
non-Markovian collision
integral which, nevertheless, retains the main
properties of the full expression~(\ref{5.22x}).
Let us approximate the correlation term~(\ref{5.21})
by its value in the quasi-equilibrium state described by
the statistical operator~(\ref{2.22}). 
In the case of weak interaction, this approximation means that
we put $f(t')\approx F(t')$. 
Then, recalling Eq.~(\ref{5.20}), we obtain
\begin{equation}
\label{5.25x}
I^C_1(t)=
{1\over\hbar^2}
\sum_{21'2'} \left|\langle 12|V|1'2'\rangle^{}_{\rm ex}\right|^2
\int_{t_0}^{t} dt'\,
\cos\left[\Delta \omega^{}_{12,1'2'}(t,t')\right]  
{\cal F}^{}_{12,1'2'}\left(\{F(t')\}\right).
\end{equation}
Now the expression~(\ref{5.22x}) takes a simpler  form
\begin{eqnarray}
\label{5.25}
& &
\hspace*{-15pt}
I^{}_1(t)= 
- {1\over\hbar^2}
\sum_{21'2'} \left|\langle 12|V|1'2'\rangle^{}_{\rm ex}\right|^2
\int_{t_0}^{t} dt'\,
\cos\left[\Delta\omega^{}_{12,1'2'}(t,t')\right]
\nonumber\\[9pt]
& &
\hspace*{160pt}
{}\times
\left[
{\cal F}^{}_{12,1'2'}\left(\{f(t')\}\right)- 
{\cal F}^{}_{12,1'2'}\left(\{F(t')\}\right)
\right].  
\end{eqnarray}
Obviously this collision integral vanishes in the quasi-equilibrium
state and, consequently, in complete equilibrium. 
Another important
point is that, in the Markovian limit, 
Eq.~(\ref{5.25}) reduces to the   
Uehling-Uhlenbeck collision integral~(\ref{5.24}) since, in this
limit, the contribution from 
${\cal F}^{}_{12,1'2'}\left(\{F(t')\}\right)$ vanishes, as 
can easily be verified with the aid of Eqs.~(\ref{5.13}) 
and~(\ref{5.8}).  
Recently, an ``improved''   
version of the Levinson collision integral was
proposed in the Green's function method 
on the basis of  approximate solution of a Dyson equation
with initial correlations~\cite{MorozovRoepke99}.
It differs from Eq.~(\ref{5.25}) 
in that the single-particle
energies did not involve the Hartree-Fock corrections and the
correlation term was approximated by 
${\cal F}^{}_{12,1'2'}\left(\{f^{({\rm eq})}\}\right)$, 
where $f^{({\rm eq})}_1$
is the distribution function in complete equilibrium.
Physically, the drug of choice for a simplified
non-Markovian collision integral is the expression~(\ref{5.25}) 
which involves effects of {\em running\/} correlations,
whereas replacing the quasi-equilibrium distribution function
$F^{}_1(t')$ by $f^{({\rm eq})}_1$ implies that the state of 
the system is
close to complete equilibrium.

\setcounter{equation}{0}

\section{Balance equations in the non-Markovian regime}
Having the explicit expression~(\ref{5.22x}) for the 
collision integral, it is of interest to  analyze in more 
detail the balance equations derived  in Section~II. 

\subsection{Energy balance. The nonequilibrium correlation energy}
We have already shown that the non-Markovian 
kinetic equation with the 
collision integral~(\ref{5.22x}) 
has the correct equilibrium solution.
Now we want to demonstrate that  the right-hand side of 
Eq.~(\ref{3.6}) can be represented as a time
derivative, i.e.,  the non-Markovian kinetic equation is
consistent with the energy conservation.    

First we will prove that any collision integral of the form
\begin{equation}
\label{6.5}
I^{}_1(t)= - \sum_{21'2'}\int_{t_0}^{t} dt'\,
\cos\left[\Delta\omega^{}_{12,1'2'}(t,t')\right]
G^{}_{12,1'2'}(t')
\end{equation}
conserves the total energy, if the function $G^{}_{12,1'2'}(t)$
has the symmetry properties
\begin{equation}
\label{6.6}
G^{}_{12,1'2'}(t)= G^{}_{21,1'2'}(t)=G^{}_{12,2'1'}(t),
\qquad
G^{}_{12,1'2'}(t)= - G^{}_{1'2',12}(t)
\end{equation}
and the function  $\Delta\omega^{}_{12,1'2'}(t,t')$ satisfies 
the conditions
\begin{equation}
\label{6.8}
\hbar\,\frac{\partial}{\partial t}
\Delta\omega^{}_{12,1'2'}(t,t')=
\Delta E^{}_{12,1'2'}(t),
\qquad
\Delta\omega^{}_{12,1'2'}(t,t)=0,
\end{equation} 
where $\Delta E^{}_{12,1'2'}(t)$ is given by Eq.~(\ref{5.16}).
The proof is as follows. Multiplying Eq.~(\ref{6.5}) by 
$E^{}_1(t)$ and then summing over the quantum numbers $1$,
we obtain 
\begin{eqnarray}
\label{6.7}
\sum_{1} E^{}_1(t)\,I^{}_1(t)&=&
- {1\over4}\sum_{121'2'} \Delta E^{}_{12,1'2'}(t)
\int_{t_0}^{t} dt'\,
\cos\left[\Delta\omega^{}_{12,1'2'}(t,t')\right]
G^{}_{12,1'2'}(t')
\nonumber\\[8pt]
{}&=&
\frac{d}{dt}
\left(
- \frac{\hbar}{4}
\sum_{121'2'}
\int_{t_0}^{t} dt'\,
\sin\left[\Delta\omega^{}_{12,1'2'}(t,t')\right]
G^{}_{12,1'2'}(t')
\right),
\end{eqnarray}
where use has been made of  Eqs.~(\ref{6.6}) and~(\ref{6.8}).
Comparison of Eqs.~(\ref{3.6}) and~(\ref{6.7}) 
shows that the non-Markovian collision integral~(\ref{6.5}) 
indeed conserves the total energy. As a by-product of the proof,
we have  the  following expression for the nonequilibrium
correlation energy:
\begin{equation}
\label{6.9}
{\cal E}^{}_{\rm corr}(t)=
{\cal E}^{}_{\rm corr}(t^{}_0)+
 \frac{\hbar}{4}
\sum_{121'2'}
\int_{t_0}^{t} dt'\,
\sin\left[\Delta\omega^{}_{12,1'2'}(t,t')\right]
G^{}_{12,1'2'}(t'),
\end{equation}
where ${\cal E}^{}_{\rm corr}(t^{}_0)$ is an initial value of the
correlation energy at $t=t^{}_0$.

Turning now to the collision integral~(\ref{5.22x})
and recalling the definition of the functionals ${\cal F}$ and 
${\cal K}$, it is easy to check that  the symmetry
conditions~(\ref{6.6}) are satisfied. Thus, the collision 
integral~(\ref{5.22x}) conserves the total energy.
In our case Eq.~(\ref{6.9}) reads
\begin{equation}
\label{6.10}
{\cal E}^{}_{\rm corr}(t)=
{\cal E}^{}_{\rm corr}(t^{}_0)
+\Delta{\cal E}^{\prime}_{\rm corr}(t)
+\Delta{\cal E}^{\prime\prime}_{\rm corr}(t),
\end{equation}
where 
\begin{eqnarray}
\label{6.11}
& &
\hspace*{-10pt}
\Delta{\cal E}^{\prime}_{\rm corr}(t)=
{1\over 4\hbar}
\sum_{121'2'} \left|\langle 12|V|1'2'\rangle^{}_{\rm ex}\right|^2
\int_{t_0}^{t} dt'\,
\sin\left[\Delta\omega^{}_{12,1'2'}(t,t')\right]
{\cal F}^{}_{12,1'2'}\left(\{f(t')\}\right),
\\[8pt]
& &
\label{6.12}
\hspace*{-10pt}
\Delta{\cal E}^{\prime\prime}_{\rm corr}(t)=
{1\over 4\hbar}
\sum_{121'2'} \left|\langle 12|V|1'2'\rangle^{}_{\rm ex}\right|^2
\int_{t_0}^{t} dt'\,
\sin\left[\Delta\omega^{}_{12,1'2'}(t,t')\right]
\nonumber\\[8pt]
& &
\hspace*{200pt}
{}\times
\frac{\beta^*(t')\,\Delta E^{}_{12,1'2'}(t')}
{\ln{\cal K}^{}_{12,1'2'}\left(\{f(t')\} \right)}
\,{\cal F}^{}_{12,1'2'}\left(\{f(t')\}\right).
\end{eqnarray}
We have separated  the time-dependent contribution to 
the correlation energy
into two parts  which have different physical interpretation. 
The term~(\ref{6.11}) can be regarded as the 
collision contribution to
the correlation energy.
An analogous term was derived  previously 
from the Levinson kinetic 
equation~\cite{Morawetz95,MorawetzKoehler98} 
and by the Green's function method~\cite{MorozovRoepke99}. 
The term~(\ref{6.12}) arises due to
collective (correlation) effects. In thermal equilibrium
these two terms cancel each other so that the correlation energy 
does not depend on time. 

It is interesting to note that the simplified non-Markovian
collision integral~(\ref{5.25}) has the form~(\ref{6.5}), 
where $G^{}_{12,1'2'}$ is the difference of two 
${\cal F}$-functionals, each of which satisfies the symmetry 
conditions~(\ref{6.6}).  We may thus conclude that 
the collision integral~(\ref{5.25})
conserves the total energy. 
The approximate correlation energy is given by
\begin{eqnarray}
\label{6.12x}
& &
\hspace*{-10pt}
{\cal E}^{}_{\rm corr}(t)=
{1\over 4\hbar}
\sum_{121'2'} \left|\langle 12|V|1'2'\rangle^{}_{\rm ex}\right|^2
\int_{t_0}^{t} dt'\,
\sin\left[\Delta\omega^{}_{12,1'2'}(t,t')\right]
\nonumber\\[8pt]
& &
\hspace*{160pt}
{}\times
\left\{
{\cal F}^{}_{12,1'2'}\left(\{f(t')\}\right)
- {\cal F}^{}_{12,1'2'}\left(\{F(t')\}\right)
\right\}.
\end{eqnarray}
Exactly  the same expression  follows from
Eq.~(\ref{6.10}) if the term~(\ref{6.12}) is replaced by 
its value in the quasi-equilibrium state.

\subsection{Non-Markovian equation for the quasi-temperature}
In general, the 
quasi-temperature evolves in time according to  Eq.~(\ref{3.16}).
Within the framework of  non-Markovian Born approximation, 
the correlation function $(\hat H,\hat f^{}_1)$ 
in this equation can be
replaced by $(\hat{\cal H}^{}_0(t),\hat f^{}_1)$ since  
the collision integral is  already of second 
order in the interaction. Then a little algebra shows that
Eq.~(\ref{3.16}) reduces to
\begin{equation}
\label{6.13}
\frac{d\beta^*(t)}{dt}=
{1\over C(t)}
\sum_{1}E^{}_1(t)\,I^{}_{1}(t).
\end{equation}
With  Eq.~(\ref{5.22x}), this is written  in the expanded 
form as
\begin{eqnarray}
\label{6.14}
& &
\hspace*{-10pt}
\frac{d\beta^*(t)}{dt}=
-\frac{1}{4\hbar^2\, C(t)}
\sum_{121'2'} \Delta E^{}_{12,1'2'}(t)
\left|\langle 12|V|1'2'\rangle^{}_{\rm ex}\right|^2
\nonumber\\[9pt]
& &
\hspace*{20pt}
{}\times
\int_{t_0}^{t} dt'\,
\cos\left[\Delta\omega^{}_{12,1'2'}(t,t')\right]
\left\{
1+
\frac{\beta^*(t')\,\Delta E^{}_{12,1'2'}(t')}
{\ln{\cal K}^{}_{12,1'2'}\left(\{f(t')\} \right)}
\right\}  
{\cal F}^{}_{12,1'2'}\left(\{f(t')\}\right).
\end{eqnarray}
The correlation function $C(t)$ is
 given by Eq.~(\ref{3.17})  
and can be calculated, in the leading
approximation, by using the statistical operator~(\ref{5.4}) 
which admits Wick's decomposition of averages. 
The Lagrange multipliers 
$\widetilde\Lambda^{}_1(t)$ can then be expressed in terms of the
one-particle distribution function  by means of
Eq.~(\ref{5.19x}).  
After some algebra, one obtains
\begin{equation}
\label{6.15}
C(t)={1\over4}
\sum_{121'2'}\left|\langle 12|V|1'2'\rangle^{}_{\rm ex}\right|^2
\bar f^{}_1(t) \bar f^{}_2(t) f^{}_{1'}(t) f^{}_{2'}(t)\,
\frac{{\cal K}^{}_{12,1'2'}\left(\{f(t)\}\right)-1}
{\ln {\cal K}^{}_{12,1'2'}\left(\{f(t)\}\right)},
\end{equation}
whence it follows that $C(t)>0$. Now the kinetic 
equation~(\ref{2.5}), together with the 
expression~(\ref{5.22x}) for
the collision integral and the evolution 
equation~(\ref{6.14}) for
the quasi-temperature, form a closed set of equations describing
non-Markovian relaxation processes in the system. 

\subsection{Entropy production in the non-Markovian regime}
Of special physical interest is the entropy equation~(\ref{3.19}). 
To second order in the interaction, the Lagrange multipliers 
$\lambda^{}_{1}(t)$ can be expressed in terms  
of the one-particle distribution functions, $f^{}_1(t)$ and
$F^{}_1(t)$,
with the aid of  Eqs.~(\ref{2.x3}), (\ref{5.19x}), and~(\ref{5.8}).
Eliminating $\widetilde\Lambda^{}_1$ and $\beta^*(E^{}_1-\mu^*)$,
we obtain 
\begin{equation}
  \label{6.1}
\lambda^{}_1(t)=
\ln\left[\frac{\bar f^{}_1(t) F^{}_1(t)}{f^{}_1(t) \bar F^{}_1(t)}
\right].  
\end{equation}
Substituting this expression, together with Eq.~(\ref{5.22}), 
into Eq.~(\ref{3.19}) and then making use of the symmetry
of the collision integral under permutations of the single-particle
quantum numbers, the entropy production takes the form
\begin{eqnarray}
\label{6.2}
& &
\hspace*{-5pt}
\frac{dS(t)}{dt}=
{1\over 4\hbar^2}
\sum_{121'2'} \left|\langle 12|V|1'2'\rangle^{}_{\rm ex}\right|^2\,
\ln\left[
\frac{{\cal K}^{}_{12,1'2'}\left(\{f(t)\}\right)}
{{\cal K}^{}_{12,1'2'}\left(\{F(t)\}\right)}
\right]
\nonumber\\[10pt]
& &
\hspace*{45pt}
{}\times
\int_{t_0}^{t} dt'\,
\cos\left[\Delta\omega^{}_{12,1'2'}(t,t')\right]
\left\{
1-
\frac{\ln{\cal K}^{}_{12,1'2'}\left(\{F(t')\} \right)}
{\ln{\cal K}^{}_{12,1'2'}\left(\{f(t')\} \right)}
\right\}  
{\cal F}^{}_{12,1'2'}\left(\{f(t')\}\right).
\end{eqnarray}
Although the right-hand side of his equation  
involves memory effects, 
the entropy production
is identically zero in thermal equilibrium, as it must
be. To understand the behavior of the entropy production in the
Markovian limit, we put $f(t')\approx f(t)$,
 $F(t')\approx F(t)$, 
and then pass to the limit $t^{}_0\to -\infty$ 
inserting the factor $\exp\{-\varepsilon(t-t')\}$.
As we have already noted, such a procedure leads to the 
appearance of
the delta function $\delta\big(\Delta E^{}_{12,1'2'}(t)/\hbar\big)$
in place of the oscillating cosine term. Due to Eq.~(\ref{5.20}),
we  may then put ${\cal K}^{}_{12,1'2'}\left(\{F(t)\}\right)=1$.
Finally, it is convenient to eliminate the ${\cal F}$-functional 
be means of the relation
\begin{equation}
\label{6.3}
{\cal F}^{}_{12,1'2'}\left(\{f\}\right)=
\bar f^{}_1 \bar f^{}_2 f^{}_{1'} f^{}_{2'}
\left({\cal K}^{}_{12,1'2'}\left(\{f\}\right)-1\right)
\end{equation}
which follows from Eqs.~(\ref{5.13}) and~(\ref{5.18}).
As a result of these manipulations, we find the entropy production in the
Markovian limit:
\begin{eqnarray}
  \label{6.4}
& &
\hspace*{-20pt}
\frac{dS(t)}{dt}=
{\pi\over 4\hbar^2}
\sum_{121'2'} \left|\langle 12|V|1'2'\rangle^{}_{\rm ex}\right|^2
\delta\left(\frac{\Delta E^{}_{12,1'2'}(t)}{\hbar}\right)  
\nonumber\\[8pt]
& &
\hspace*{70pt}
{}\times\left[{\cal K}^{}_{12,1'2'}\left(\{f(t)\}\right) -1 \right]
\ln{\cal K}^{}_{12,1'2'}\left(\{f(t)\}\right)
\left(\bar f^{}_1 \bar f^{}_2 f^{}_{1'} f^{}_{2'}\right)^{}_t.
\end{eqnarray}      
It depends only on the one-particle distribution function and
is positive, since $(x-1)\ln x\geq 0$ for $x>0$. 
The result~(\ref{6.4}) agrees with the well-known
expression for the entropy production in a weakly interacting
quantum system described by a Markovian kinetic 
equation (see, e.g.,~\cite{AkhiezerPeletminski81}).   

\setcounter{equation}{0}

\section{Generalization to spatially non-homogeneous
systems}
Here we shall briefly touch upon the extension of the
foregoing treatment to spatially
non-homogeneous systems. 
In such cases, many-particle correlations
associated with the energy conservation can be incorporated by
taking mean energy density 
${\cal E}({\bf r},t)=\langle \hat H({\bf r})\rangle^t$ 
as a state variable, together with the one-particle Wigner
function 
$f^{}_{\alpha}({\bf r},{\bf p},t)=
\langle \hat f^{}_{\alpha}({\bf r},{\bf p})\rangle^{t}$.
The energy density operator, $\hat H({\bf r})$, is defined
through the relation
\begin{equation}  
\label{D.1}
\hat H= \int d{\bf r}\, \hat H({\bf r}),
\end{equation}  
and the operator corresponding to the Wigner function is given by 
\begin{equation}
\label{D.2}
\hat f^{}_{\alpha}({\bf r},{\bf p})=
\int d{\bf x}\,{\rm e}^{-{\bf p}\cdot{\bf r}/\hbar}\,
\psi^{\dagger}_{\alpha}({\bf r}-{\bf x}/2)\,
\psi^{}_{\alpha}({\bf r}+{\bf x}/2), 
\end{equation}
where $\psi^{}_{\alpha}({\bf r})$ and 
$\psi^{\dagger}_{\alpha}({\bf r})$ are the second-quantized
field operators; the Greek indices $\alpha=(i,\sigma)$ specify
the type of particles $(i)$ and the spin state $(\sigma)$.
Now the relevant statistical operator takes the form 
[cf.~Eq.~(\ref{2.13})]
\begin{eqnarray}
\label{D.3} 
& &
\hspace*{-20pt}
\varrho^{}_{\rm rel}(t)=
{1\over Z^{}_{\rm rel}(t)}\,
\exp\left\{
-\int d{\bf r}\,\beta^{*}({\bf r},t)
\left[
\hat H({\bf r}) - \sum_{i} \mu^{*}_{i}({\bf r},t)\,
\hat n^{}_{i}({\bf r})\right]
\right.
\nonumber\\[6pt]
& &
\hspace*{120pt}
\left.
{}-\sum_{\alpha} \int \frac{d{\bf p}}{(2\pi\hbar)^{3}}\,
\lambda^{}_{\alpha}({\bf r},{\bf p},t)\,
\hat f^{}_{\alpha}({\bf r},{\bf p})
\right\},
\end{eqnarray}
where 
$\hat n^{}_{i}({\bf r})=
\sum_{\sigma} \psi^{\dagger}_{\alpha}({\bf r})
 \psi^{}_{\alpha}({\bf r})$ 
are the particle-number density operators for the species.
The local inverse quasi-temperature $\beta^{*}({\bf r},t)$ and
the Lagrange multipliers 
$\lambda^{}_{\alpha}({\bf r},{\bf p},t)$ are to be determined
from the local equations of state
\begin{equation}
\label{D.4}
{\cal E}({\bf r},t)=
{\rm Tr}\left\{ \hat H({\bf r})\,\varrho^{}_{\rm rel}(t)  
\right\},
\qquad
f^{}_{\alpha}({\bf r},{\bf p})=
{\rm Tr}\left\{
\hat f^{}_{\alpha}({\bf r},{\bf p})\,\varrho^{}_{\rm rel}(t)
\right\}.
\end{equation}
Just as in the spatially homogeneous case,
the local quasi-chemical potentials 
$\mu^{*}_{i}({\bf r},t)$ in Eq.~(\ref{D.3}) can be eliminated by
re-defining the Lagrange multipliers 
$\lambda^{}_{\alpha}({\bf r},{\bf p},t)$. 
Another possible way is to define the $\mu^{*}_{i}({\bf r},t)$
in the local-equilibrium state described by the statistical
operator 
\begin{eqnarray}
\label{D.5}
\varrho^{}_{\rm loc}(t)=
{1\over Z^{}_{\rm loc}(t)}\,
\exp\left\{
-\int d{\bf r}\,\beta^{*}({\bf r},t)\left[
\hat H({\bf r}) -  \sum_{i} \mu^{*}_{i}({\bf r},t)\,
\hat n^{}_{i}({\bf r})
\right]
\right\}.
\end{eqnarray}    
Then the quasi-chemical potentials are to be determined from the
local equations of state
\begin{equation}
\label{D.6}
n^{}_{i}({\bf r},t)=
{\rm Tr}\left\{\hat n^{}_{i}({\bf r})\,\varrho^{}_{\rm loc}(t)
\right\},
\end{equation}
where $n^{}_{i}({\bf r},t)=
\langle \hat n^{}_{i}({\bf r})\rangle^{t}$ are
the mean particle-number densities for the species.
The latter definition of the quasi-chemical potentials  
is typical for the hydrodynamic description of transport 
processes. 

In the spatially non-homogeneous case, the basic evolution
equations are the kinetic equation for the Wigner function
\begin{equation}
\label{D.7}
\frac{\partial}{\partial t}\,
f^{}_{\alpha}({\bf r},{\bf p},t)=
\frac{1}{i\hbar}\,
\big\langle [\hat f^{}_{\alpha}({\bf r},{\bf p}),\hat H]
\big\rangle^{t}
\end{equation} 
and the local conservation law for the mean energy density
\begin{equation}
\label{D.8}
\frac{\partial}{\partial t}\,{\cal E}({\bf r},t)=
\frac{1}{i\hbar}\,
\big\langle [\hat H({\bf r}),\hat H]
\big\rangle^{t}=
- \nabla\cdot \langle \hat{\bf J}({\bf r})\rangle^{t},
\end{equation}
where $\hat{\bf J}({\bf r})$ is the energy flux operator.
To calculate the averages appearing in Eqs.~(\ref{D.7}) 
and~(\ref{D.8}), the nonequilibrium statistical operator has to
be found as a functional of the state variables.
This can be done by solving Eq.~(\ref{4.5}) in some approximation 
with the relevant statistical operator given by Eq.~(\ref{D.3}).   
If, for instance, the interaction term in the Hamiltonian can be
regarded as a small perturbation, then Eq.~(\ref{4.5}) can be
solved by an iterative method which is similar to the
procedure used in Section~III. There are, however, some new
features which are specific to spatially non-homogeneous systems.
First, now the average on the right-hand side of 
Eq.~(\ref{D.7}), 
when calculated with the relevant statistical operator, is not
zero and the kinetic equation for the Wigner function
has the form
\begin{equation}    
\label{D.9}  
\frac{\partial}{\partial t}\,
f^{}_{\alpha}({\bf r},{\bf p},t) +
D^{}_{\alpha}({\bf r},{\bf p},t)=
I^{}_{\alpha}({\bf r},{\bf p},t),
\end{equation}
where 
\begin{equation}
\label{D.10}
D^{}_{\alpha}({\bf r},{\bf p},t)=
\frac{i}{\hbar}\,
\big\langle [\hat f^{}_{\alpha}({\bf r},{\bf p}),\hat H]
\big\rangle^{t}_{\rm rel}
\end{equation}
is a generalized drift term which involves the correlation
contributions~\cite{MorozovRoepke95}. The collision
integrals $I^{}_{\alpha}({\bf r},{\bf p},t)$ 
are determined by the integral terms in Eq.~(\ref{4.5}).
The second important point is that, 
in the spatially non-homogeneous
case, the right-hand side of Eq.~(\ref{4.5}) contains terms 
with gradients of the Lagrange multipliers, 
$\beta^{*}({\bf r},t)$, $\mu^{*}_{i}({\bf r},t)$, and
$\lambda^{*}_{\alpha}({\bf r},{\bf p},t)$.
Finally, the collision integrals 
$I^{}_{\alpha}({\bf r},{\bf p},t)$ are, in general, 
non-local functionals of the Wigner function.
If the state parameters vary little over the range
of the interaction potential and the mean de Broglie 
wavelength, which is typical for real situations, then 
the non-locality effects can be incorporated by using
an expansion of averages in Eqs.~(\ref{D.7}) 
and~(\ref{D.8}) in powers of spatial 
gradients. Within this scheme, the evolution equations derived
in this paper may be interpreted as transport equations in the
local approximation, where all the gradient terms  
are neglected.   
 
\section{Conclusions and outlook}  
We now summarize the main implications of the above analysis. 
The first point is that  a non-Markovian kinetic equation 
conserves the total energy and has the correct equilibrium 
solution only if   the collision  and correlation effects are 
incorporated in a self-consistent way, 
for instance, 
within the same approximation in the interaction 
between particles.  
We have also seen that it is precisely the  
interplay between correlations  and collisions
that is responsible for the long-time 
behavior of the non-Markovian collision integral.

Another important feature of the outlined approach is that
the kinetic equation for the one-particle distribution function 
is supplemented by the 
equation for the quasi-temperature, Eq.~(\ref{6.13}),
which describes the ``slow'' evolution of the system. 
This representation for non-Markovian kinetics differs from 
standard schemes, say the Green's function 
method~\cite{KadanoffBaym62,Danielewicz84a,Botermans90},
where the goal is to derive a closed kinetic equation, 
so that the collision and correlation effects have to be 
described in terms of single-particle quantities. 
The inclusion of  long-lived  many-particle
correlations into the Green's function formalism is a 
rather difficult
problem and there are only first steps  in this 
way~\cite{Kukharenko82,MorawetzRoepke95,Bornath96}. 
On the other hand, the Green's function technique 
provides a powerful 
tool for calculating the collision contribution to a 
kinetic equation beyond the Born approximation. 
Thus an interesting point would be a unification of the 
Green's function method and the density operator method to
develop the self-consistent non-Markovian quantum kinetic theory 
involving the correlation effects and the quasiparticle damping.

It is significant that the approach outlined 
in this paper is 
non-perturbative in external fields. 
If the external field does not
directly affect interactions between particles, then  the
nonequilibrium statistical operator will have the form~(\ref{4.7}),
but the evolution operator $U^{}_0(t,t')$ will now involve
the field effects which can be taken into account exactly
(not in terms of perturbation theory).  
It should also be noted that, in the presence of  
an external field, the 
energy of the system is not conserved and  the trivial
equation $d{\cal E}/dt=0$ has therefore to be replaced by a 
balance equation
including the work produced by the field.

\end{document}